\documentstyle[12pt,epsf]{article}
\newcommand{\bq}{\begin{equation}}
\newcommand{\eq}{\end{equation}}
\newcommand{\ra}{\rightarrow}
\newcommand{\ov}{\overline}
\newcommand{\al}{\langle}
\newcommand{\ar}{\rangle}
\newcommand{\J}{J/\Psi}
\newcommand{\Ps}{\Psi^\prime}

\begin{document}

\begin{flushright}
{BINP 99-54}\\
hep-ph/9906387
\end{flushright}

\begin{center}{\bf $J/\Psi$  THEORY REVIEW}\\ or \\
 {\bf FROM\,\,\,\, $J$\,\,\,\, TO \,\,\,\, $\Psi$}\end{center}
\vskip 1cm
\begin{center} {\bf Victor Chernyak}\end{center}
\begin{center}
Budker Institute of Nuclear Physics,\,\, Novosibirsk,\,\, Russia 
\end{center}
\vspace{0.5cm}
\begin{center} Talk given at the International Workshop \end{center}
\begin{center}
{\bf "{\boldmath $e^+e^-$} collisions from {\boldmath $\phi$} to \boldmath{
$\J$"}} \end{center} 
\begin{center}1-5 March 1999, Novosibirsk\end{center}
\vspace{0.5cm}
\begin{center} { CONTENT} \end{center}

1) Hard exclusive processes in QCD: main characteristic properties

2) Experimental and theoretical status of $J/\Psi$ and $\Psi^\prime$ decays\\

\begin{center}
I.\,\,\,{\bf Asymptotics of exclusive amplitudes in QCD}.
\end{center}

Exclusive processes are those in which not only the initial, but also the 
final state is completely specified.  
\footnote{
In contrast, the inclusive process does not specifies the final state and is
a sum of all possible exclusive final states. The well known example is the 
deep inelastic scattering $e+p\ra e^\prime+X$, for which the cross section
includes summation over all possible final states $X$. }  Therefore, all 
experimentalists dealt with exclusive processes, but may be not
all of them recognized this.

In what follows, we will be interested in two types of exclusive processes: 
a) reactions $e^+e^-\ra \gamma^{*}\ra H_1\,H_2$, where $H_1$ and $H_2$ are
two definite hadrons, mesons or baryons; these reactions determine 
electromagnetic form factors of hadrons, both elastic and transition ones;
b) decays of the heavy quarkonium into two definite light hadrons, say
$\J\ra \rho\pi,\,\J\ra \pi\pi$, etc.

Starting from the original papers [1,\,2], the general theory of hard $-$
i.e. those at large energies and with large momentum transfers $-$ exclusive 
processes in QCD was well developed, see i.e. \cite{ch3}. In particular, it 
has been shown that, due to the asymptotic freedom of QCD, the leading power 
behaviour is determined by the connected Born diagrams, while loop corrections
result only in an additional slow logarithmic evolution, similarly to the 
deep inelastic scattering. 

For instance, the leading power behaviour of the electromagnetic form factor 
at $Q^2\ra \infty$ can be obtained as follows. In the Breit frame, the 
initial and final hadrons move along the z-axis with large momenta, 
$p_{1z}=(Q/2),\, p_{2z}= (-Q/2)$. So, the quarks inside hadrons also have 
large longitudinal momenta: $k_z^i=x_i\,p_{1z}$ and $q_z^i=y_i\,p_{2z}$ for 
the initial and final quarks respectively, where $x_i,\,y_i$ are the hadron 
momentum fractions carried by the i-th quark,\, $0\leq x_i,\,y_i\leq 1.$ 
At the same time, all quark transverse momenta remain small, $k_{x,y}^i\sim
q_{x,y}^i\sim \Lambda_{QCD}$. Thus, to obtain the leading contribution to the 
matrix element of the electromagnetic current $\langle H_2(p_2)|J_{\mu}|H_1
(p_1)\rangle$, it is legitimate to neglect the quark transverse momenta 
at all, as well as their binding energy inside hadrons which is also $\sim
\Lambda_{QCD}$, and to replace the initial and final hadron states by a set
of free collinear quarks, see figs.1,2.
\footnote 
{ On figures: the thick line is the charm quark, the thin line is
the light quark, the dashed line is the gluon, the wavy line is the photon.}
One obtains then for the meson, see fig.1
: $\al M(p_2)|J_{\mu}|M(p_1) \ar \sim \al {\bar d}(y_2 p_2),\,u(y_1
p_2)|J_{\mu}|\,{\bar d}(x_2 p_1),\,u(x_1 p_1)\ar
\sim (\sqrt {E_2})^2 (1/Q)(1/Q^2)(\sqrt {E_1})^2\sim (1/Q)$. 
The above factors originate from: a) two final quark spinors, each behaving 
as $\sim \sqrt {E_2}\sim \sqrt Q$; b) the quark propagator $\sim (1/Q)$; c) 
the gluon propagator $\sim (1/Q^2)$; d) two initial quark spinors, each 
behaving as $\sim \sqrt {E_1}\sim \sqrt Q$. For the pion, for instance,
because $\al \pi^{+}(p_2)|J_{\mu}|\pi^{+}(p_1) \ar =(p_1+p_2)_{\mu} F_{\pi}
(Q^2)$ and $p_1\sim p_2 \sim Q,$ it follows from the above expression 
that the pion form factor behaves as: $F_{\pi}(Q^2)\sim (1/Q^2)$. In the same 
way, one obtains for the nucleon matrix element, fig.2: $\al N(p_2)|
J_{\mu}|N(p_1)\ar \sim (\sqrt{E_2})^3\,(1/Q)^2\,(1/Q^2)^2\,(\sqrt {E_1})^3
\sim (1/Q^3)$. Because $\al N(p_2)|J_{\mu}|N(p_1)\ar \sim {\ov N}_2\gamma_
{\mu}N_1\cdot F_N(Q^2)$, where $N_1$ and ${\ov N}_2$ are the initial and
final nucleon spinors (each behaving as $\sim \sqrt Q$) and $F_{N}(Q^2)$ is
the nucleon form factor, it follows from the above that 
$F_{N}(Q^2)\sim 1/Q^4.$

In fact, proceeding in the above described simplest way one obtains the 
highest possible power behaviour of form factors and, similarly, of
any other hard exclusive amplitude. The real behaviour of 
amplitudes can be additionally power suppressed depending on the
quantum numbers of the initial and final hadrons. There are two main 
selection rules for these additional suppressions \cite{ch1}. 

a). The first one is relevant for the higher helicity hadron 
states: $|\lambda|=|S_z| > 1$ for mesons and
$|\lambda| > 3/2$ for baryons. Such meson helicities can not be made from
the quark spin projections, $S_{z1},\,S_{z2}=(\pm 1/2)$, which means 
the quarks have necessarily a nonzero angular momentum projection
$L_z=[\lambda-S_{z1}-S_{z2}]\neq 0$ in such a state
\footnote{
 Considering here the valence wave function 
component, i.e. those with the minimal number of constituents.
}, 
and its wave function $\Phi_i$ contains the factor $\sim \exp\{iL_z\phi\}$.   

The total transition amplitude is a product of the initial and final hadron 
wave functions, $\Phi_i$ and $\Phi_f^{*}$, with the hard kernel $T$ (which is 
a product of all intermediate quark and gluon propagators), integrated over 
the quark relative momenta inside the initial and final hadrons. These
integrations include in particular $\int_0^{2\pi}d\phi$ (and similarly for
other hadrons). So, with respect to the angle $\phi$, the transition 
amplitude contains the factor: 
$$\int_0^{2\pi}d\phi\exp\{iL_z\phi\}T(Q,x_j,{\bf k}_{\perp}, \dots)\,\,.$$ 

The $\phi$-dependence appears in $T$ through its dependence on the
quark transverse momenta in the state $\Phi_i$: ${k}_{\perp }^{\pm}=(k_{x}
\pm i k_{y})=|{\bf k}_{\perp}|\exp \{\pm i \phi\}$. The hard kernel can be 
thought as being decomposed into a series in powers of $k_{\perp}^+$ and $k_
{\perp}^-$. Then, the above integration separates out the term $\sim \exp\{-
iL_z\phi\}$ from this series, while all terms $\sim \exp\{-iL_z^\prime\phi\}$
with $L_z^\prime\neq L_z$ will give zero after integration over $\phi$. 

The crucial point is that "the hard kernel is hard", i.e. intermediate
quark and gluon propagators have virtualities $\sim Q^2$, due to large 
logitudinal
momentum transfers $\sim Q$. On the other hand, the typical quark transverse
momenta inside the hadron are small, $|{\bf k}_{\perp}|\sim \Lambda_{QCD}$. 
So, the quark transverse momenta give only small power corrections in the
hard kernel. The expansion of the hard kernel looks schematically as
follows: $T=Q^{-n}[1+(k_{\perp}^{\pm}/Q)+(k_{\perp}^{\pm}/Q)^2+\dots]$. 
Therefore, the needed term extracted from the hard kernel will
have the additional suppression $\sim (k^{-}_{\perp}/Q)^{L_z}\sim (\Lambda_
{QCD}/Q)^{L_z}\exp\{-iL_z\phi\}$, 
in comparison with the leading behaviour of the hard kernel.
\footnote{This is for $L_z>0$\,;\,\,
for $L_z<0$ the corresponding term is $\sim (k^{+}_{\perp}/Q)^{(-L_z)}$.}

Thus, the final result is that for those hadron states which require the
wave function components with $L_z\neq 0$, each unit of $|L_z|$ will result
in the additional suppression factor $\sim (1/Q)$ in the amplitude. Therefore,
{\it the leading contribution to the hard exclusive amplitudes originates only
from the hadron wave functions components with $L_z=0$, in which the
hadron helicity is a sum of spin projections of its constituent quarks}:
$\lambda$={\bf n}{\bf J}=$\sum_i$ ({\bf n}{\bf S}$_i$). Let us emphasize 
that {\it it is $L_z\neq 0$, not the angular momentum $L\neq 
0$ by itself}, which leads to the amplitude suppression. 
So, {\it the hard exclusive amplitudes for those mesons and baryons
which are the P,\,D, etc. $-$ states in the quark model are not power 
suppressed in comparison with the S-state hadrons}. The reason is that the
light quarks are relativistic inside the hadrons and, besides, $Q$ is much 
larger than hadron masses, while the power suppression of amplitudes for the 
$L\neq 0$ states is right in the completely nonrelativistic situation only.  

b) The second selection rule is much more evident. It can be formulated as
follows: {\it the quark helicities are conserved inside the hard kernel}.
This originates from the fact that in QCD (like QED) the helicity of the 
energetic quark is conserved in perturbation theory, because the 
quark-gluon interaction is vectorlike. For this reason, the perturbative 
helicity flip amplitude is $\sim (m_q/E)$, in comparison with the helicity 
conserving one. Because the current (i.e. entering the QCD Lagrangean) masses
of the {\it $u-,\, d-$} and {\it $s-$} quarks are small, $m_u\simeq 4\,MeV,\, 
m_d\simeq 7\, MeV,\, m_s\simeq 150\,MeV$, the corrections to the leading term
from such helicity flip contributions are small for the strange quark, 
$\sim (m_s/Q)$, and tiny for the $u-$ and $d-$ quarks.

It seems at first sight that much larger helicity flip contributions 
can originate from the
dynamically generated (as a result of nonperturbative interactions leading to
spontaneous breaking of the axial symmetry) constituent quark masses, which
are at low quark virtualities: 
$M^{const}_{u,d}\simeq 350\, MeV,\,\,M_s=m_s+ M^{const}_s\simeq 500\, MeV.$ 
Really, this is not the case. The reason is that the constituent quark
masses $M^{const}_i$ are "soft", unlike the current quark masses $m_i$
which are "hard". This means that the current quark mass is only weakly 
(logarithmically) dependent on the quark virtuality, while the constituent 
quark mass, generated by soft nonperturbative interactions, tends quickly to 
zero at high virtualities: $M_i^{const}(k^2)\sim \al 0|{\ov \psi}\psi|0\ar/
k^2\sim \Lambda_{QCD}^3/k^2$. So, $M_i^{const}(k^2)$ give really only small 
power corrections inside the hard kernel, because the quark virtualities are 
large here: $k^2_i\sim Q^2$. The same considerations are applicable to all 
other nonperturbative effects, not only to those connected with generation
of $M^{const}$. Because all nonperturbative interactions are "soft", i.e.
their effects are power suppressed at high virtualities $\sim (\Lambda_{QCD}
^2/k^2)^n,$ they all produce only power corrections inside the hard kernel. 

The quark helicity shows its spin projection onto its momentum. But the 
directions of the quark and the hadron momenta do not coincide exactly 
because quarks inside hadrons have nonzero transverse momenta, $k_{\perp}^i$. 
Let us recall once more, however, that the quark longitudinal (i.e. along 
the hadron momentum) momentum is proportional to the hadron momentum, 
$k_{\parallel}^i=x_i|{\bf p}|\sim Q$,
while their transverse momenta remain small, $k_{\perp}^i\sim \Lambda_{QCD}$. 
So, the angle between the quark momentum and the hadron momentum is small,
$\theta\sim (|k_{\perp}|/k_{\parallel})\sim (\Lambda_{QCD}/Q)$. Thus, up to 
these power corrections, the conservation of the quark helicity is equivalent to
conservation of its spin projection onto the momentum of the hadron it 
belongs. 

Taken alone, this does not lead to immediate consequencies for hadron
helicities because the hadron helicity differs in general from the sum of
quark spin projections when $L_z\neq 0$. But being combined with the point 
"a" above that $L_z=0$ for the leading contributions, the net result is 
{\it a conservation of hadron helicities in the leading contributions to all 
hard exclusive amplitudes.} These include in particular all diagonal and
transition form factors, the heavy quarkonium decays into light hadrons, etc. 

All the above considerations and selection rules are 
summarized by a simple formula for the asymptotic power behaviour 
of any form factor \cite{ch1}:
\bq 
\al p_2, J_2,\lambda_2|J_{\lambda}|p_1, J_1, \lambda_1\ar \sim
\left ( \frac{1}{Q}\right )^{ (n^{min}_1+n^{min}_2-3)+|\lambda_2-
\lambda_1|}\,\,\,\,. 
\eq
Here: $p_i,\,J_i,\,\lambda_i$ are the initial and final hadron momenta, 
spins and helicities
respectively;\, $\lambda$ is the photon helicity and $\lambda=(\lambda_1+
\lambda_2)$ in the Breit frame;\, $n^{min}_i$ are the minimal possible 
numbers of constituents inside the given hadron: $n^{min}=2$ for mesons
and $n^{min}=3$ for baryons. The hadron helicity conservation is clearly 
seen from eq.(1). Moreover, it determines the behaviour of all 
nonleading form factors.

In agreement with the explanations given above, {\it the asymptotic behaviour 
is independent of hadron spins, parities etc., and only helicities are the
relevant quantum numbers}.
Besides, increasing the number of constituents results in additional
suppression, so that the nonvalence components (i.e. those containing
additional gluons or quark-antiquark pairs) of the hadron wave functions
give power suppressed corrections.
\footnote{
In fact, the above given explanations were somewhat simplified. For instance,
the wave function of the meson state with $\lambda=2$ contains not only the 
two quark component with $S_{z1}=S_{z2}=1/2,\,L_z=1$, but also, say, the 
nonvalence component with the additional gluon, such that: $S_{z1}=S_{z2}=
1/2,\,S_{z3}=1,\,L_{zi}=0$. It is not difficult to see that the final result
remains the same. The form factor gains $\sim Q$ due to $L_z=0$ here instead 
of $L_z=1$ in the two quark component, but losses $\sim 1/Q$ because 
$n_1=3$ now.
}  
\begin{center}
II.\,\,\,{\bf  $\J$ and $\Ps$ decays}
\end{center}

As was described above, there are simple rules allowing one to obtain the
leading power term of any exclusive amplitude at $Q\ra \infty$, and a large
number of concrete calculations has been performed, see i.e. \cite{ch3}.
As for the loop corrections to the Born contributions, they are amenable to 
standard perturbation theory calculations and result in a slow additional
logarithmic evolution with increasing Q, see [1-3].

Up to now, however, there remains the main unsolved problem in all practical
applications $-$ to calculate (or to estimate reliably, at least) the power
corrections to the leading terms. These power corrections, which are
negligble in the formal limit $Q\ra \infty$, may be of real importance when
comparing the leading term calculations with the present data which have
typically $Q=2-4\,GeV$. It seems at first sight that $Q\simeq 3\,GeV$, which
corresponds to charmonium decays, is sufficiently large in comparison with
the scale of power corrections which is typically $\simeq \Lambda_{QCD}\simeq 
350\,MeV$. The matter is, however, that the role of power corrections is
determined actually not by the ratio $(\Lambda_{QCD}/Q)$, but rather by $(
\Lambda_{QCD}/Q_{eff})$, with $Q_{eff}\ll Q$. To illustrate, let us consider 
the $\J\ra K^*K$ decay. In the $\J$ rest frame
the momentum of each light meson is $ |{\bf p}|=1.37\,GeV$. It is
shared however between two its constituent quarks, so that the longitudinal
momentum of each quark is roughly: $k_{\parallel}\simeq |{\bf p}|/2\simeq 
680\,MeV$, which is only a factor two larger than its transverse momentum 
$|{\bf k_{\perp}}|\simeq 350\,MeV$. So, the power corrections to the leading 
term due to $|{\bf k_{\perp}}|\neq 0$ can be $\simeq 50\%$ in the amplitude. 

The size of power corrections can be estimated also in a pure kinematical 
way. Say, the
$\J\ra b_1(1235)\pi$ decay amplitude contains the term $\sim (e_{\mu}^
{\lambda_o}\epsilon_{\mu}^{\lambda_1})$, where $e^{\lambda_o}$ and $\epsilon
^{\lambda_1}$ are the polarization vectors of $\J$ and $b_1$ respectively.
At large values of the $b_1$-momentum: $\epsilon_{\mu}^{\lambda_1=0}=(p_{\mu}
/M_{b_1})+O(M_{b_1}/|{\bf p}|)$, while $\epsilon_{\mu}^{|\lambda_1|=1}
=O(1)$. So, the production of the $|b_1^{|\lambda_1|=1}\pi\ar$-state will be
power suppressed, $\sim (M_{b_1}/|{\bf p}|)$, in comparison 
with those of $|b_1^{\lambda_1=0}\pi\ar$. Really however the suppression is 
absent as $(M_{b_1}/|{\bf p}|)=0.95$ in the $\J\ra b_1\pi$ decay. On the
other hand, such kinematical power corrections can be taken under control,
unlike the "dynamical" power corrections mentioned in the preceding 
paragraph.

On the whole, the heavier are final mesons the
worse is a situation, and it is even worse for two baryon decays as the
baryon momentum is shared between its three quarks. 

These simple estimates show that there are no serious reasons to 
expect that formally leading terms will really dominate the charmonium 
decay amplitudes. 

In particular, there are no reasons to expect that the above described 
helicity selection rules will be actually operative here, as the factor 
$\simeq (1/2)$ "suppression" (see above) can be easily overhelmed by others 
numerical factors. For instance, the formally leading contribution to $\J\ra 
VT$ and $\J\ra AP$ decays gives the fig.3 diagram 
\footnote{
$\Psi_i$ means $\J$ or $\Ps$, in this diagram and in all others.}
in which the meson wave functions
contain the minimal number of constituents, and with no helicity flips.
Due to a loop however, it contains the additional smallness $\simeq ({\ov
\alpha}_s/\pi)\simeq 0.1$, so that the figs.4-6 diagrams which are formally
suppresed $\sim (\Lambda_{QCD}/Q_{eff})$ due to the nonvalence (i.e. three
particle) wave functions, give really larger contributions.   

Thus, it looks somewhat strange that a number of authors insist that those 
amplitudes which are helicity suppressed (or contain the nonvalence 
wave function components which give the same effect) in the formal limit 
$Q\ra \infty$, will be actually heavily suppressed in charmonium decays. 
The experiment shows that this is not the case. 
For instance, the $(\J\ra \omega\pi)$-decay amplitude is formally helicity 
suppressed while $(\J\ra \pi^+\pi^-)$ is not, but\, \cite{PDG} : Br$(\J\ra 
\omega\pi)/$Br$(\J\ra\pi^+\pi^-)\simeq 3$\,. \\

A large number of various $\Ps$ decays has been measured by the BES
Collaboration during last time [5-7], and these very 
interesting results were presented at this workshop by Prof. Stephen L. 
Olsen \cite{Ol}. So, it is a great challenge for theory to
understand and explain these data.

When comparing the $\J$ and $\Ps$ decays into light hadrons, it will be wrong 
to compare the branchings by itself. The reason is that $\Ps$, unlike $\J$,
decays mainly to lower charmonium states. Besides, the values of the $\Ps$
and $\J$ wave functions at the origin are different. To avoide both these
differences, one rather has to compare the ratios: Br$(\Ps\ra X)$/Br$(\Ps\ra 
e^+e^-)$ and  Br$(\J\ra X)$/Br$(\J\ra e^+e^-)$.
In other words, one has to rescale the $\Ps$-branchings by the factor\,
\cite{PDG}: Br$(\Ps\ra e^+e^-)$/Br$(\J\ra e^+e^-)=0.14\,$.  
Actually, this is not the only scale factor, as the mass of $\Ps$ is
noticeably higher than those of $\J:\, (M^2_{\Ps}/M^2_{\Psi})=1.4$,
and the exclusive branchings have a high power dependence on the initial 
mass,\, $\sim (1/M)^{n_{eff}}$,\, see tables. So, it seems reasonable to 
separate out this dependence which is, besides, quite different for 
different decay channels. On the whole, we have to compare the 
"reduced" decay amplitudes, $A$ and $A^\prime$ for $\J$ and $\Ps$ 
respectively, from which both the above described rescaling factors are
separated out. In the tables 1-5 given below I have tried to recalculate 
these reduced amplitudes from the experimental data (taking into account 
also the phase space corrections).

Now, the natural expectation is that, in a "normal situation", these reduced 
amplitudes $A^\prime$ and $A$ will be close to each other, so that $R=|A^
\prime/A|$ will be close to unity. As it is seen
from the tables 1-5, the situation is not "normal". Some decay channels are
strongly suppressed ($R\ll 1$), while other ones are significantly enhanced
($R > 1$). 

The most famous is the $"\rho\pi$-puzzle", i.e. a very small value of
Br$(\Ps\ra \rho\pi)$/Br$(\J\ra \rho\pi)$, see table 1.
A number of speculative explanations have been proposed, including even so
exotic as a significant admixture of the charm component, $({\ov C}C)$, in 
the $\rho$-meson wave function \cite{bk}. Most of speculations are based on 
the idea that $\Ps\ra \rho\pi$ is "naturally small" because the decay
amplitude is helicity suppressed, while $\J\ra \rho\pi$ is "abnormally
large". So, the efforts were concentrated on searching the sources of this
enhancement: especially introduced nearby gluonium resonance \cite{res}, or
a large admixture of the colored component in the $\J$-wave
function, unlike the $\Ps$ one \cite{braa}, etc. 

As was pointed out above, the helicity suppression is not really a strong 
effect in the charmonium region as its typical value is only $\simeq 1/2$ 
here, and it is easily overhelmed by other numerical factors. So, the whole
above idea does not look very appealing. Moreover, the
(hopefully) main contributions to $\J\ra \rho\pi$ have been directly 
calculated long ago in \cite{zzc} (see also \cite{ch3}, ch. 9.1). 
These originate from the figs.4-5 diagrams and give: Br$(\J\ra \rho\pi)
\simeq 1\%$. This shows clearly that the experimental value: Br$(\J\ra
\rho\pi)=(1.27\pm 0.09)\%$ \cite{PDG} is natural and there is no need 
for an additional enhancement. 

So, we come naturally to the idea that it is not $\J\ra \rho\pi$
which is enhanced, but rather $\Ps\ra \rho\pi$ which is suppressed. What may 
be the reason ? Looking at tables 1-5 one sees that the underlying
mechanism is highly nontrivial and is of dynamical nature as it
operates selectively, suppressing some channels and enhancing other ones.
The possible explanation has been proposed long ago in \cite{ch3} (ch. 8.4).
The idea is that $\Ps$, unlike $\J$, is really a highly excited state, $-$ 
it is close even to the ${\ov D}D$-threshold\,! So, it looks natural that
there is a large admixture of the nonvalence $-$ i.e. those containing the 
additional gluon or the light quark-antiquark pair $-$ components in its wave
function. If so, its decays into various channels can differ significantly
from those of $\J$, for which the nonvalence component is expected to be less 
significant because it is the lowest state. Unfortunately, this idea also 
remains a pure speculation, as no concrete calculations were performed up to 
now. So, by necessity, our discussion below will be more qualitative than 
quantitative.

{\bf 2.1\,\,  VP - decays} (see table 1).\quad As for the strong amplitudes, 
the main diagrams 
are expected to be those shown on figs.4-6. For $\Ps$, according to the above
described idea, the diagrams like those shown on fig.7 (and many others 
similar) are expected to be also of importance. Let us emphasize that, even 
in the formal limit $Q=M_{\Ps}\ra \infty$, these nonvalence contributions are 
not power suppressed here in comparison with the valence ones. Indeed, two 
out of three gluon propagators in the diagram on fig.7 (denoted by open 
circles) are semihard, i.e. their virtuality is parametrically only $k_i^2
\sim p_o Q$, rather than $\sim Q^2$, where $p_o$ is the light quark bound 
state momentum in the $\Ps$-rest frame. This gains the factor $\sim (Q^2/p_o^
2)$. Besides, the $\omega$ and pion wave functions are two-particle here and 
both can have the leading twist. This gains the factor $\sim (Q/\Lambda_{QCD})
$, in comparison with the figs.4-6 contributions where one out of two mesons 
has the nonleading twist three-particle wave function. The whole
gained factor $\sim (Q^3/p_o^2\Lambda_{QCD})$ compensates for the additional 
smallness $\sim (p_o^3/Q^3)$ due to the additionall $({\ov q}q)$-pair
in the 4-particle component of the $\Ps$-wave function. 

So, the proposed here
explanation of the $\rho-\pi$-puzzle is that the valence and nonvalence
strong contributions interfere destructively in this channel and, as a matter
of case, cancel to a large extent in the total $\Ps\ra \rho\pi$ strong 
amplitude, while the role of nonvalence contributions is much less 
significant in $\J\ra\rho\pi$. From this viewpoint, there is no deep reason 
for the experimentally observed very strong suppression of $\Ps\ra\rho\pi$, 
this is a result of a casual cancelation. 

{\bf 2.2\,\, VT and AP - decays} (see table 2).\quad These decays are the 
leading ones in the 
formal limit $Q\ra \infty$. However, these formally leading contributions 
originate from the loop diagram, fig.3, and thus contain the additional loop
smallnes $\simeq ({\ov \alpha}_s/\pi)\simeq 0.1$. Therefore, they are small 
really in comparison with contributions of the figs.4-6 diagrams which are 
expected to be dominant in the $\J$-decays. 
Thus, the prediction is that all $\J\ra (VP,\,VT,\,AP)$  decays receive 
main contributions from the figs.4-6 diagrams and will 
be of comparable strength, and this is the case, see tables 1-2. As for the
$\Ps\ra (VT,\,AP)$  decays, it is seen from the table 2 that the sign of the
interference of valence and nonvalence contributions is opposite in these
two channels, so that the $VT$ - decays are suppressed while the $AP$ - ones 
are enhanced.

{\bf 2.3}\,\, {\boldmath ${\ov B}B$}\, {\bf - decays} (see table 3).\quad 
It is seen from the table 3 
that all baryon decays are enhanced at $\Ps$. This may result from the 
nonvalence contributions like those shown in fig.8 (and others similar), 
interfering 
constructively with the valence contributions from the diagram on fig.9. 

{\bf 2.4\,\, PP and VV - decays} (see table 4).\quad There is a number of 
"puzzles". First, the 
$\pi^+\pi^-$-mode which is electromagnetic is (it seems, the error bars are 
large) strongly enhanced at $\Ps$, in comparison with that at $\J$. If this
decay mode were solely due to the pion form factor $F_{\pi}(Q^2)$, fig.10, 
this will imply $F_{\pi}(M^2_{\Ps})\simeq 2\,F_{\pi}(M^2_{\J})$. But this is  
impossible because $F_{\pi}(Q^2)$ decreases with $Q^2$, like $\sim 1/Q^2$. 
So, this is a strong indication that the nonvalence contributions (like those
shown on fig.11, and many others similar) are of great importance here. This 
is not the whole
story however, as Br$(\J\ra \pi^+\pi^-)$ is too large by itself.  If it were
solely due to $F_{\pi}(Q^2)$, then it will result in: $F_{\pi}(M_{\J}^2)
\simeq 1\,GeV^2/M_{\J}^2$, while most of theoretical estimates do not exceed
$\simeq (0.5-0.6)\,GeV^2/M_{\J}^2.$ So, we come to a conclusion that, in this 
channel, the nonvalence contributions are of importance even for $\J$.

If really important, the nonvalence contributions like those shown on the 
fig.11 diagram will have also another impact. The matter is that their SU(3) 
flavour structure is different from those of the fig.10 diagram, and this 
possibility was ignored in all phenomenological descriptions of $\J\ra PP$ - 
decays (and similarly for the nonvalence electromagnetic contributions in 
other decay channels). So, such contributions can influence, in particular, 
the conclusions about $\sim 90^o$ relative phases of the strong and 
electromagnetic contributions to the decay amplitudes. 

{\bf 2.5}\,\, {\boldmath $(\J,\,\Ps\ra \omega\pi^o)$}\, {\bf - decays} 
(see table 1).\quad This 
channel is electromagnetic (i.e. it needs photon) and is very interesting. 
Suppose first that it is solely due to the $\gamma\omega\pi$ - form factor
$F_{\omega\pi}(Q^2)$, fig.10. This last is helicity suppressed and has 
the asympotic behaviour: $\sim (1/Q^4)$. 
\footnote{
$F_{\omega\pi}(Q^2)$ is defined by: $\al \omega_{\lambda}(p_1)\pi^o(p_2)|J_
{\mu}|0\ar =\epsilon_{\mu\nu\sigma\tau}p_{1}^{\nu}p_{2}^{\sigma}e_{\lambda}
^{\tau}\cdot F_{\omega\pi}(Q^2)$. The helicity of $\omega$ is $|\lambda|=1$ 
here. From eq.(1), this matrix element 
behaves as $\sim (1/Q^2)$, resulting in $F_{\omega
\pi}(Q^2)\sim (1/Q^4)$, because $p_1\sim p_2\sim Q,\, e_{\lambda}\sim 1$.}
The data at $Q^2\leq 
5\,GeV^2$ are sufficiently well described by a sum of $\rho(770)$ and $\rho
^{\prime}(1460)$ contributions. Being not too close to these resonances, we 
can neglect the widths, and it will be given then by a simple expression:
\bq
F_{\omega\pi}(Q^2)=F_{\omega\pi}(0)\frac{m_{\rho}^2\,M_{\rho^\prime}^2}
{(m_{\rho}^2-Q^2)(M_{\rho^{\prime}}^2-Q^2)}\,\,;\quad F_{\omega\pi}(0)=2.3\,
GeV^{-1}.
\eq
But the eq.(2) fails to describe the $\J\ra\omega\pi$ and $\Ps\ra\omega\pi$ 
decays, as this will require much larger value of $F_{\omega\pi}(Q^2=M_
{\J}^2)$ and especially of $F_{\omega\pi}(Q^2=M_{\Ps}^2)$. Moreover,
the reaction $e^+e^-\ra \omega\pi^o$, which by definition determines $F_
{\omega\pi}(Q^2)$, has been measured recently by the BES Collaboration. The
results \cite{Ol} for $F_{\omega\pi}(Q^2)$ are presented on fig.12, together
with previous data at smaller $Q^2$ [13,\,14]. 
The curve is from eq.(2). It is seen
that the new data agree with eq.(2), while both $\J$ and $\Ps$ lie well above.
This shows unambiguously that both $(\J\ra \omega\pi)$ and $(\Ps\ra \omega
\pi)$ - decay amplitudes contain, besides $F_{\omega\pi}$, additional 
significant contributions, and these last are larger for $\Ps$ than for $\J$.
All this agrees with the additional nonvalence contributions, like those 
shown on the fig.13 diagram (and many others similar; here too these 
nonvalence contributions are not power suppressed in comparison with the 
valence ones, even in the formal limit $Q\ra \infty$).

{\bf 2.6}\,\, {\boldmath $(\J,\,\Ps\ra \gamma\pi^o)$}\, {\bf - decays} 
(see table 5). \quad There are two
main contributions to the $(\J\ra \gamma\pi^o)$ - amplitude: the VDM one, 
fig.14, and through the intermediate photon, fig.15. They were calculated in
\cite{zzc} (see also \cite{ch3}, ch.5.4) with the result: Br$(\J\ra \gamma
\pi^o)\simeq 4\cdot 10^{-5}$, in agreement with data. The relative 
contributions to the decay amplitude from these diagrams 
are here: (fig.14):(fig.15)$\simeq (1.6\,:\,1.0)\,.$ When going from $\J$ to 
$\Ps$, the radical difference emerges from the strong suppression of the VDM 
contribution for $\Ps$ due to nonvalence contributions, while these last do
not influence the amplitude with the intermediate photon. Thus, the 
relative contributions to the decay amplitude become now 
(using the BES data for $\Ps\ra \rho\pi$, see table 1): (fig.14):(fig.15)
$\simeq (0.3\,:\,1.0)$, and Br$(\Ps\ra \gamma\pi^o)$ will be $\simeq 1\cdot 
10^{-6}$.  

\begin{center}{\bf Conclusions}\end{center}

A large number of various exclusive $\Ps$-decays has 
been measured by the BES Collaboration during last time. These results
are very interesting not only for understanding the properties of $\Ps$ by 
itself. Comparison of $\Ps$ and $\J$ decays helps to elucidate the
properties of both of them, as well as properties of strong interaction at
these energies. 

In particular, these measurements revealed that, fortunately or 
unfortunately, the situation is much more complicated than the naive 
expectations based on the dominance of formally leading (at $Q\ra \infty$) 
terms. This especially concerns $\Ps$ which is a highly excited state close
to the ${\bar D}D$-threshold. 

The present status of theory is such that, it 
seems, there is some understanding of the $\J$ and $\Ps$ - decays on the 
qualitative and sometimes on the semiquantitative level, but a lot of job 
remains to be done to have a really quantitative description of these decays. 

As it is clear from the above considerations, it will be of great interest
and of great help for theory if our colleagues experimentalists will be able 
to measure some distinguished electromagnetic processes somewhere in the
region $ 8\,GeV^2 < Q^2 < 16\,GeV^2$, but out of $\J$ and $\Ps$
(like the recent measurement of $e^+e^-\ra \omega\pi$ by the BES 
Collaboration, see fig.12). They are: $e^+e^-\ra \{\pi^+\pi^-,\, K^+K^-,\, 
\rho\eta,\, \rho\eta^{\prime},\, \gamma \pi,\, \gamma\eta,\, \gamma\eta^
{\prime}\}\,$. These measurements will determine the corresponding form 
factors which are of great interest by itself and will be useful for 
elucidating the electromagnetic contributions into the $\J$ and $\Ps$ - 
decays. Besides, it will be very useful to improve the poor accuracy of 
$\Ps\ra \pi^+\pi^-$ and $\Ps\ra K^+K^-$ and to measure 
$\Ps\ra {\bar K}^o K^o$.

\begin{center}{\bf Acknowlegments}\end{center}

I am deeply grateful to L.M. Barkov, A.E. Bondar, B.I. Khazin and S.I.
Serednyakov for useful discussions and critical remarks, and to N.I. Root 
for preparing the figure 12. 

This work is supported in part by the grant INTAS 96-155.

\newpage

{\small
\begin{center}{\bf Table 1}\,: $\quad (J/\Psi,\,\, \Psi^{\prime})\ra VP\,;
\quad n_{eff}=6,\,\,\, k=3$ \end{center}

\begin{center}
\begin{tabular}{|c|c|c|c|c|}

\hline
{\bf CHANNEL} & {\bf Br} $(\psi^{\prime}\ra VP)\cdot 10^4$ &
{\bf Br} $(\psi\ra VP)\cdot 10^3$ & $\Delta \%$ & $R=|A^{\prime}/A|$  \\
\hline \hline
$\rho\,\pi$ & $< 0.83\,$ PDG & $12.7\pm 0.9$ & 5.0 & $< 0.36\,$ PDG \\
 & $< 0.28\,$ BES & PDG & & $< 0.21\,$ BES\\
\hline \hline
$K^+ K^{* -}$ & $< 0.54\,$ PDG & $ 5.0\pm 0.4$ & 5.0 & $< 0.46\,$ PDG\\
+c.c. & $< 0.30\,$ BES & PDG & & $< 0.35\,$ BES\\
\hline \hline
${\ov K}^o K^{* o}$ & $0.81\pm 0.24\pm 0.16 $ & $4.2\pm 0.4$ & 5.0 &
$0.62\pm 0.14$\\
+c.c. & BES & PDG & & BES\\
\hline \hline
$\omega\eta $ & $< 0.26\,$ & $1.58\pm 0.16$ & 5.0 & $< 0.58$\\
 & BES & PDG & & BES\\
\hline\hline
$\omega \eta\prime$ & $0.76\pm 0.36\pm 0.15$ & $0.17\pm 0.02$ & 7.4 & 
$2.5\pm 0.8$\\
 & BES & PDG & & BES\\
\hline\hline
$\phi\eta$ & $0.35\pm 0.19\pm 0.07$ & $0.65\pm 0.07$ & 5.7 & $1.0\pm 0.3$\\
 & BES & PDG & & BES\\
\hline\hline
$\phi\eta^{\prime}$ & $< 0.75 $ & $0.33\pm 0.04$ & 6.5 & $< 1.9$ \\
 & BES & PDG & & BES\\
\hline\hline\hline
$\omega\pi^o$ & $0.38\pm 0.17\pm 0.11$ & $0.42\pm 0.06$ & 5.0 & 
$1.35\pm 0.35$\\
 (electromagnetic) & BES & PDG & & BES\\
\hline\hline
$\rho \eta$ & $0.21\pm 0.11\pm 0.05$ & $0.19\pm 0.02$ & 5.0 & $1.5\pm 0.5$\\
(electromagnetic) & BES & PDG & & BES\\
\hline\hline
$\rho \eta\prime $ & $< 0.3$ & $0.10\pm 0.02$ & 7.4 & $< 2.0$\\
(electromagnetic) & BES & PDG & & BES\\
\hline
\end{tabular}
\end{center}
Here and in all other tables, the total rescaling factor $\Delta$ is:
$$\Delta=\frac{Br(\psi^{\prime}\ra {\ov e}e)}{Br(\psi\ra {\ov e}e)}\,\,\,
\cdot\left ( \frac{M_{\psi}}{M_{\psi^{\prime}}}\right )^{n_{eff}}\cdot
\left (\frac {M_{\psi}\cdot p^{\prime}}{M_{\psi^{\prime}}\cdot p}
\right )^{k}\,\,\,,$$
where the meaning of the first factor was explained in the text,
the second factor accounts for dependence of
decay amplitudes on the mass of the decaying charmonium state, and the last
one is the phase space correction ( $p$ and $p^{\prime}$ are the c.m.s. 
momenta of final particles in the $\J\ra X$ and $\Ps\ra X$ decays).

The ratio of the "reduced " amplitudes is: 
$$R=\left |\frac{A^\prime}{A}\right |=\left \{ \frac{1}{\Delta}\,\frac{Br 
(\Ps\ra X)}{Br (\J\ra X)}\right \}^{1/2},$$
where $A$ and $A^\prime$ are the "reduced" amplitudes of the $\J$ and $\Ps
\ra X$ decays. 
}
\newpage
{\small
\begin{center}
{\bf Table 2}\,: $(J/\Psi,\,\,\Psi^{\prime})\ra VT,\,\,AP\,; \quad
n_{eff}=6,\,\,\,k=1$
\end{center}

\begin{center}
\begin{tabular}{|c|c|c|c|c|}
\hline
{\bf CHANNEL} & {\bf Br} $(\psi^{\prime}\ra X)\cdot 10^4$ &
{\bf Br} $(\psi\ra X)\cdot 10^3$ & $\Delta \%$ & $R=|A^{\prime}/A|$  \\
\hline \hline
$\omega f_2(1270) $ & $< 1.7\,$ & $4.3\pm 0.6$ & 5.6 & $< 0.85\pm 0.06$\\
 & BES & PDG & & BES\\
\hline\hline
$\rho a_2(1320)$ & $< 2.3$ & $10.9\pm 2.2$ & 5.6 & $< 0.61\pm 0.06$\\
 & BES & PDG & & BES\\
\hline\hline
$\phi f_2^{\prime}(1525)$ & $< 0.45 $ & $1.2\pm 0.2 $ & 6.6 & $ < 0.75\pm 
0.10 $\\
 & BES & PDG & & BES\\
\hline \hline
$K^{*,o}{\ov K}_2^o(1430)$ & $< 1.2$ & $6.7\pm 2.6$ & 6.0 & $< 0.55\pm 
0.10$\\
+c.c. & BES & PDG & & BES\\
\hline\hline\hline
$b_1^+(1235)\,\pi^-$ & $5.3\pm 0.8\pm 0.8$ & $3.0\pm 0.5$ & 5.3 & 
$1.82\pm 0.25$  \\
 & BES & PDG & & BES\\
\hline\hline
$K_1^{+}(1270) K^-$ & $10.0\pm 1.8\pm1.8$ & $< 2.9$ & 5.4 & 
$> 2.5\pm 0.3$\\
+c.c. & BES & BES & & BES\\
\hline \hline
$K^+_1(1400) K^-$ & $< 2.9 $ & $3.8\pm 0.8\pm 0.5$ & 5.5 &
$< 1.2\pm 0.15$\\
+c.c. & BES & BES & & BES\\
\hline
\end{tabular}
\end{center}
}
{\small
\begin{center}
{\bf Table 3}\,: $(J/\Psi,\,\,\Psi^{\prime})\ra {\ov B}B\,; \quad
n_{eff}=8,\,\,\,k=1$
\end{center}

\begin{center}
\begin{tabular}{|c|c|c|c|c|}
\hline
{\bf CHANNEL} & {\bf Br} $(\psi^{\prime}\ra X)\cdot 10^4$ &
{\bf Br} $(\psi\ra X)\cdot 10^3$ & $\Delta \%$ & $R=|A^{\prime}/A|$  \\
\hline \hline
${\ov p}p$ & $1.9\pm 0.5$ & $2.14\pm 0.10$ & 3.8 & 
$1.5\pm 0.2$  \\
 & PDG & PDG & & PDG\\
\hline \hline
${\ov \Lambda}\Lambda$ & $2.11\pm 0.23\pm 0.26$ & $1.27\pm 0.17$ & 4.0 & 
$2.0\pm 0.2$\\
 & BES & PDG & & BES\\
\hline \hline
${\ov \Sigma^o}\Sigma^o $ & $0.94\pm 0.30\pm 0.38$ & $1.8\pm 0.4$ & 4.1 &
$1.35\pm 0.35$\\
 & BES & PDG & & BES\\
\hline \hline
${\ov \Xi}\Xi $ & $0.83\pm 0.28\pm 0.12$ & $1.35\pm 0.14$ & 4.6 & 
$ 1.0\pm 0.2$\\
 & BES & PDG & & BES\\
\hline\hline
${\ov \Delta}^{--}\Delta^{++}$ & $0.89\pm 0.10\pm 0.24$ & $1.10\pm 0.29$ & 
4.2 & $ 1.4\pm 0.3$\\
 & BES & PDG & & BES\\
\hline
\end{tabular}
\end{center}
}
\newpage
{\small 
\begin{center}{\bf Table 4}\,: $\quad (J/\Psi,\,\, \Psi^{\prime})\ra PP,\,VV;
\quad n_{eff}=4,\,\,\, k=3$

\end{center}

\begin{center}
\begin{tabular}{|c|c|c|c|c|}
\hline

{\bf CHANNEL} & {\bf Br} $(\psi^{\prime}\ra X)\cdot 10^4$ &
{\bf Br} $(\psi\ra X)\cdot 10^3$ & $\Delta \%$ & $R=|A^{\prime}/A|$  \\
\hline \hline
$\pi^+\,\pi^-$ & $0.8\pm 0.5$ & $0.15\pm 0.02$ & 7.0 & $2.8\pm 0.9\,$ \\
(electromagnetic) & PDG & PDG & & PDG\\
\hline \hline
$K^+ K^-$ & $1.0\pm 0.7$ & $ 0.24\pm 0.03$ & 7.3 & $2.4\pm 0.9\,$\\
 & PDG & PDG & & PDG\\
\hline \hline
${\ov K}^o K^o$ &  & $0.11\pm 0.02$ & 7.3 & \\
 & & PDG & &\\
\hline \hline
${\ov K}^{*,o}K^{*,o}$ & $0.45\pm 0.25\pm 0.07$ & $< 0.5$ & 8.6 & 
$> 1.0\pm 0.3$\\
 & BES & PDG & & BES\\
\hline

\end{tabular}
\end{center}
}
{\small
\begin{center}{\bf Table 5}\,: \end{center}

\begin{center}
$\quad (J/\Psi,\,\, \Psi^{\prime})\ra \gamma P,\, \gamma T;
\quad n_{eff}^{(\gamma \eta, \gamma\eta^\prime)}=4,\, k^{(\gamma P)}=3\,;\,\,
n_{eff}^{(\gamma T)}=2,\, k^{(\gamma T)}=1$
\end{center}

\begin{center}
\begin{tabular}{|c|c|c|c|c|}
\hline

{\bf CHANNEL} & {\bf Br} $(\psi^{\prime}\ra X)\cdot 10^4$ &
{\bf Br} $(\psi\ra X)\cdot 10^3$ & $\Delta \%$ & $R=|A^{\prime}/A|$  \\
\hline \hline
$\gamma \eta$ & $0.53\pm 0.31\pm 0.08$ & $0.86\pm 0.08$ & 7.1 & 
$0.93\pm 0.30\,$ \\
 & BES & PDG & & BES\\
\hline \hline
$\gamma \eta^{\prime}(958)$ & $1.54\pm 0.31\pm 0.23$ & $ 4.31\pm 0.3$ & 7.7 &
$0.68\pm 0.08$\\
 & BES & PDG & & BES\\
\hline \hline \hline
$\gamma f_2(1270)$ & $3.0\pm 1.1\pm 1.1$ & $1.38\pm 0.14$ & 10.0 & 
$1.5\pm 0.4$  \\
 & BES & PDG & & BES \\
\hline \hline
$\gamma f_2^{\prime}(1525)$ &  & $0.47\pm 0.06$ & 11.0 & \\ 
&  & PDG & & \\
\hline\hline\hline
$\gamma\pi^o$ & $\simeq 1\cdot 10^{-2}$ & $0.039\pm 0.013$ &  & \\
 & prediction & PDG & &\\
\hline
\end{tabular}
\end{center}
}
\begin{figure}[htb]
\vspace{-1cm}
\epsfysize=24cm
\epsffile{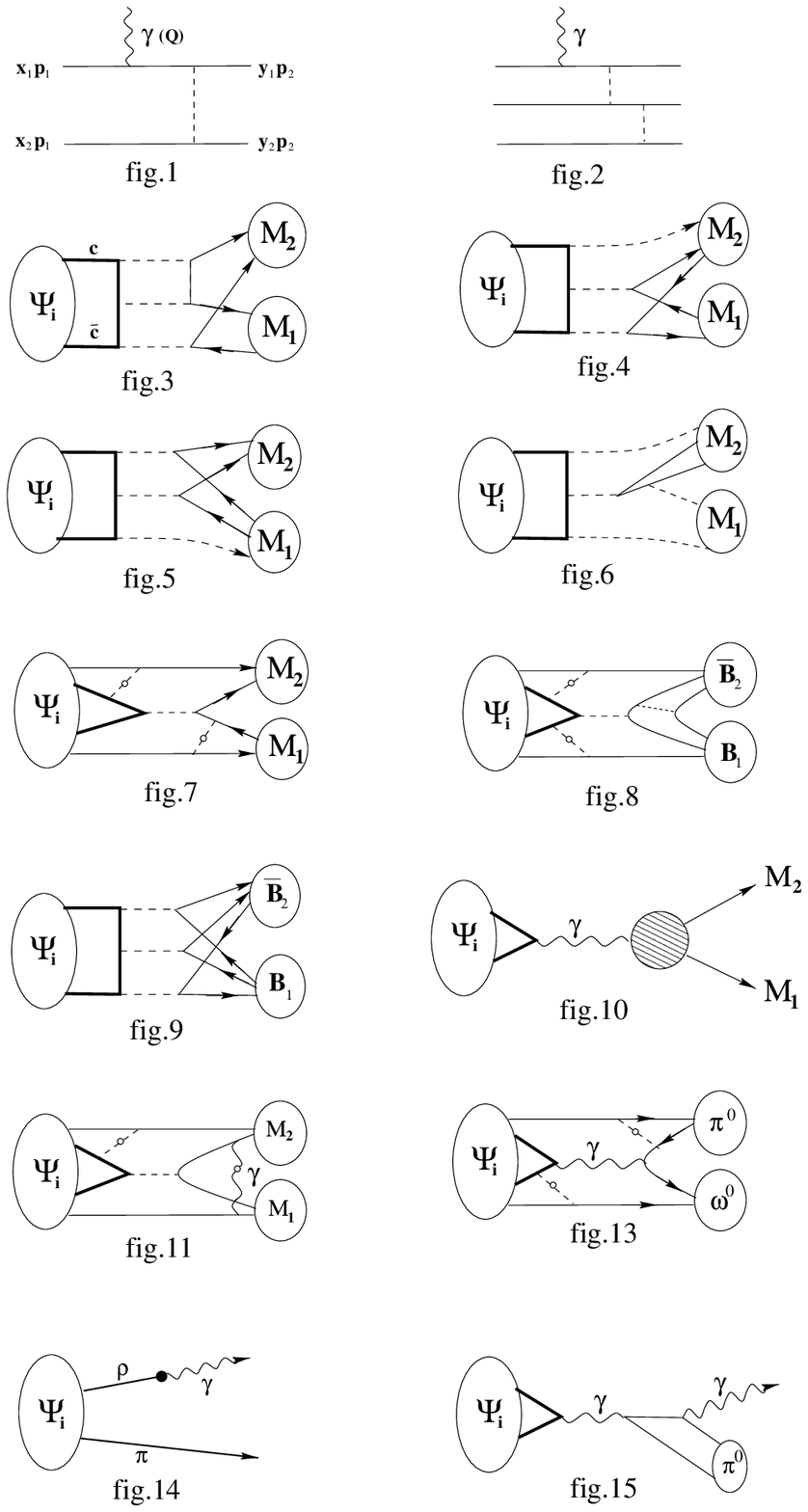}
\end{figure}

\begin{figure}[htb]
\hspace*{-2cm}
\epsfxsize=18cm
\epsffile{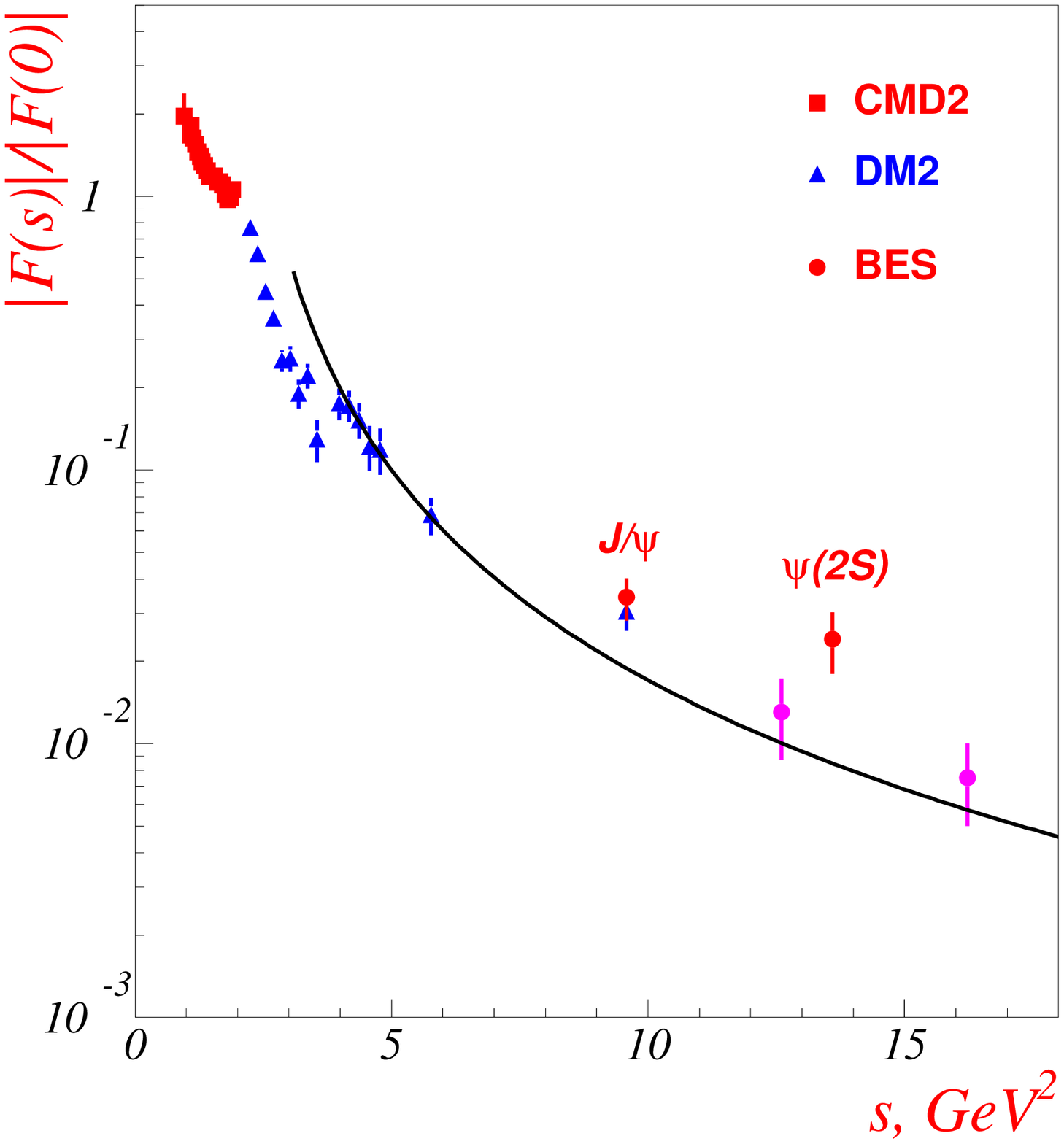}
\addtocounter{figure}{11}

\caption{\LARGE{
The curve: $\frac{F_{\omega \pi}(s)}{F_{\omega \pi}(0)}=
\frac{m_{\rho}^2\,M_{\rho \prime}^2}{(m_{\rho}^2-s)(M_{\rho \prime}^2-s)
}\,\,$;}
\newline
\normalsize
$\,\,\,\,m_{\rho}$=0.77\,GeV,\,\,\,$M_{\rho \prime}$=1.46\,GeV 
}
\end{figure}


\newpage
\begin{thebibliography}{25}
\bibitem{ch1}
V.L. Chernyak and A.R. Zhitnitsky, JETP Lett., {\bf 25} (1977) 510;\\
Sov J. Nucl. Phys., {\bf 31} (1980) 544.
\bibitem{ch2}
V.L. Chernyak, V.G. Serbo and A.R. Zhitnitsky, JETP Lett., \\
{\bf 26} (1977) 594;\,\,\, Sov. J. Nucl. Phys., {\bf 31} (1980) 552.
\bibitem{ch3}
V.L. Chernyak and A.R. Zhitnitsky, Phys. Rep. {\bf 112} (1984) 173.
\bibitem{PDG}
Particle Data Group, Europ. Phys. J., {\bf C 3} (1998) 1.
\bibitem{rad}
BES Collaboration, Phys.Rev., {\bf D 58} (1998) 097101.
\bibitem{VT}
BES Collaboration, Phys.Rev.Lett., {\bf 81} (1998) 5080.
\bibitem{AP}
BES Collaboration, hep-ex/9901022.
\bibitem{Ol}
S.L. Olsen (BES Collaboration), \\
Puzzles in hadronic physics around 3 GeV, These Proceedings.
\bibitem{bk}
S.J. Brodsky and M. Karliner, Phys.Rev.Lett., {\bf 78} (1997) 4682.
\bibitem{res}
W.S. Hou and A.Soni, Pys.Rev.Lett., {\bf 50} (1983) 569;\\
S.J.Brodsky, G.P.Lepage and S.F.Tuan, Phys.Rev.Lett., {\bf 59}(1987)621.
\bibitem{braa}
Yu-Qi Chen and E. Braaten, Phys.Rev.Lett., {\bf 80} (1998) 5060.
\bibitem{zzc}
A.R. Zhitnitsky, I.R. Zhitnitsky and V.L. Chernyak,\\
Yad. Fiz., {\bf 41} (1985) 199; and ref. [3], ch. 9.1, ch. 5.4.
\bibitem{kmd}
R.R. Akhmetshin et al., CMD-2 Collaboration\\
Preprint BINP 98-63, Novosibirsk 1998, hep-ex/9904024.
\bibitem{dm2}
D. Bisello et al., DM2-Collaboration, \\
Nucl. Phys. (Proc. Suppl.) {\bf B 21} (1991) 11.
\end{thebibliography}
\end{document}